\documentclass{aa}  
\usepackage{natbib}
\usepackage{epsf}
\usepackage{graphicx}

\begin{document}
\def\aap{A\&A}
\def\apj{ApJ}
\def\apjl{ApJL}
\def\mnras{MNRAS}
\def\aj{AJ}
\def\nat{Nature}
\def\aaps{A\&A Supp.}
\def\me{m_\e}
\def\lesssim{\mathrel{\hbox{\rlap{\hbox{\lower4pt\hbox{$\sim$}}}\hbox{$<$}}}}
\def\gtrsim{\mathrel{\hbox{\rlap{\hbox{\lower4pt\hbox{$\sim$}}}\hbox{$>$}}}}

   \title{Cluster scaling relations from cosmological hydrodynamic
   simulations in dark energy dominated universe}

   \author{N.~Aghanim
          \inst{1}
          \and
          A.~C.~da Silva\inst{1,2}
	  \and
	  N.~J.~Nunes \inst{3}
          }

   \offprints{}

   \institute{IAS, CNRS \& Universit\'e Paris Sud, B\^atiment 121, 91405
     Orsay, France \\
              \email{nabila.aghanim@ias.u-psud.fr}
         \and
             Centro de Astrofisica, Universidade do Porto, Rua das
             Estrelas, 4150-762 Porto, Portugal \\
             \email{asilva@astro.up.pt}
         \and
             Department of Applied Mathematics and Theoretical
              Physics, Wilberforce Road, Cambridge, CB3 0WA, UK \\
             \email{nunes@damtp.cam.ac.uk}
             }

   \date{}

 
  \abstract {Clusters are potentially powerful tools for cosmology
provided their observed properties such as the Sunyaev-Zel'dovich (SZ)
or X-ray signals can be translated into physical quantities like mass
and temperature.  Scaling relations are the appropriate mean to
perform this translation. It is therefore, important to understand
their evolution and their modifications with respect to the physics
and to the underlying cosmology.}  {In this spirit, we investigate the
effect of dark energy on the X-ray and SZ scaling relations. The study
is based on the first hydro-simulations of cluster formation for
diferent models of dark energy.
We present results for four dark energy models which differ
from each other by their equations of state parameter, $w$. Namely, we
use a cosmological constant model $w=-1$ (as a reference), a perfect
fluid with constant equation of state parameter $w=-0.8$ and one with
$w = -1.2$ and a scalar field model (or quintessence) with varying
$w$.}  {We generate $N$-body/hydrodynamic simulations that include
radiative cooling with the public version of the Hydra code, modified
to consider an arbitrary dark energy component. We produce cluster
catalogues for the four models and derive the associated X-ray and SZ
scaling relations.} { We find that dark energy has little effect on
scaling laws making it safe to use the $\Lambda$CDM scalings for
conversion of observed quantities into temperature and masses.} {}


\titlerunning{Cluster scaling relations from cosmological hydrodynamic 
simulations of dark energy}
\authorrunning{Aghanim et al.}

\maketitle

\section{Introduction}

In order to explain the current acceleration of the universe
\cite{riess1998, perlmutter1999} in the context of the theory of
General Relativity, it is general procedure to introduce a new form of
gravitational component with negative pressure -- dark energy.
Various candidates such as a cosmological constant or a quintessence
field have been proposed.  These models are characterised by their
equation of state parameter and constrained by either using the
observation of background quantities or the growth of cosmic
structures. An alternative way to explain the acceleration of the
universe is to allow for modifications of gravity. Many classes of
models exist, for instance, a light scalar field coupled to matter
leads to models of extended quintessence and more generally to
scalar-tensor type theories. Further possibilities were studied in
the context of braneworld models. Testing the Poisson equation on
large scales may be a way to distinguish between all these alternative
scenarios.

Regardless of its nature, dark energy as a dominant component, plays a
role in structure formation and thus modifies the number of formed
structures. The evolution of linear perturbations in the presence of a
scalar field like quintessence and the effects on the abundance of
collapsed structures and its dependence with redshift were widely
explored and suggested as a tool to constrain the nature of dark
energy and its evolution, e.g. \cite{haiman2001, weller2001,
weinberg2003, battye2003, wang2004, mohr2004}.  The properties of
collapsed halos (density contrast and virial radius) depend strongly
on the shape of the potential, the initial conditions, the time
evolution of the dark energy equation of state and on its ability to
collapse at the structure scale, e.g. \cite{nunes2004}.  One can
additionally investigate the effects of dark energy on the growth rate
of structure and consequently study how dark energy affects the
abundance of collapsed halos in dark energy models,
e.g. \cite{nunes2006, manera2006}.

Cluster number counts can potentially discriminate between dark
energy models, in particular, the evolution of their equation of
state. However, a large number of clusters with known redshifts from
SZ, X-ray or optical surveys are needed, 
e.g. \cite{bartelmann05}. Future Sunyaev-Zel'dovich (SZ) as well as
X-rays observations will provide us with high quality data making
galaxy clusters an efficient and powerful cosmological tool. Cluster
physics, however, is complex. The presence of substructures, the
possible contamination by radio and IR sources, the still imperfect
knowledge of the relation between the halo mass and the clusters
observed properties induce degeneracies between cluster physics and
cosmological models. Scaling relations are key quantities in
observational cosmology as they relate the observables in X-rays and
SZ to the cluster properties, namely, masses and temperatures. The
latter are then used in cluster number counts to constrain the
cosmological models.

Due to their importance in translating observations to physical
quantities, but most of all in probing the cluster formation, scaling
relations have raised considerable attention.  Simple models of
formation of virialised systems such as clusters predict that they
exhibit self-similar behaviours \cite{kaiser1986}, see also
\cite{ascasibar2006}. 
In the self-similar model, gravitational infall drives shock heating
of the intra-cluster medium (ICM) and establishes the gas properties
such that they scale with the halo mass giving rise to the scaling
relations. However, it is now clear that additional physics is
required to provide a more complete picture of the cluster formation
and evolution, and to explain the deviations between observations and
the predictions based on self-similar scaling as shown mostly by X-ray
observations
\cite{edge91,allen98,markevitch98,nevalainen2000,finoguenov2001,ettori2004,henry2004,arnaud2005,rasia2005,balogh2006,maughan2006,morandi2007}.
As the complexity of the physical description increases due to the
additional gas physics (galactic winds and/or quasar outflows,
radiative cooling, preheating) the use of numerical simulations
appears the best option to compare predictions and observations and
examine the role of those new ingredients in explaining the departures
form self similarity
(e.g. \cite[]{evrard96,bryan98,bialek2001,thomas2001,babul2002,
voit2002,borgani2004,rowley2004,muanwong2006,kay2007}).  The ICM can
additionally be probed by the Sunyaev-Zel'dovich (SZ) effect, inverse
Compton scattering of cosmic microwave background (CMB) photons off
high energy electrons. The magnitude of the SZ effect is determined by
the integrated gas pressure along the line-of-sight, called the
Compton parameter, $y$.  Unlike the X-ray surface brightness, the SZ
effect is not subject to cosmological dimming and can be used out to
high redshift which makes SZ scaling relations particularly
interesting and attractive tests. Nevertheless, SZ measurements,
although gaining in quality and quantity, are at the moment still not
sufficient to be fully used and this explains why investigations of SZ
scaling relations are still limited from the observational point of
view \cite[]{cooray99, benson2004, mccarthy2003, morandi2007} as well
as from the numerical point of view
\cite[]{dasilva2004,molt2005,bonaldi2007}.

Scaling laws relate the observed properties in X-rays and SZ to the
cluster properties, namely, masses and temperatures which are then
used to construct mass functions and number counts utilised to
constrain the cosmological models. Understanding the possible biases
in the scaling relations is thus essential for the use of clusters as
cosmological probes. Previous studies, based on numerical simulations,
have focused on the effects of additional gas physics in the scaling
laws. In the present study, we explore the effects of the cosmological
model on the scaling properties of galaxy clusters and their
evolutions. We perform hydrodynamic numerical simulation of cluster
formation and evolution assuming a simple radiative cooling model,  
rather than a more complete gas model,
which allow us to single out only dark energy properties.
We then investigate weather the X-ray and SZ scaling laws derived for
different dark energy models depart from those obtained in the
standard cosmological constant model (the $\Lambda$CDM model) taken as
our reference. In the next section we briefly present the dark energy
models used for the first numerical simulation of cluster formation
with hydrodynamics in dark energy dominated universes. We describe the
simulation code and the procedure used to construct the X-ray and SZ
cluster catalogues. In Sect. \ref{sec:scal}, we present the X-ray and
SZ scaling laws studied in the article together with the fitting
procedure. Our results and conclusions are summarised in
Sect. \ref{sec:res} and \ref{sec:conc} respectively.

\section{Dark-energy simulations}

\subsection{Simulation models}

Numerical N-body simulations including a dark energy component were
performed by several groups to complement the analytical computations
of structure formation in presence of dark energy, and to the study the
effects of dark energy at the structure level. 
All studies on galaxy clusters, were essentially dedicated to study
dark matter halo shapes and mass functions in different models of dark
energy \cite{Linder:2003dr, Lokas:2003cj, Klypin:2003ug, dolag:2004,
  Kuhlen:2004rw}.  The overall picture that has emerged from these
studies is that halo mass functions are well approximated by the
Jenkins mass function \cite{jenkins:2001} and halo core densities, or
concentrations, are sensitive to the mean cosmological density at halo
formation and therefore depend on the underlying dark-energy
model. These findings have led to investigations of the effect of dark
energy on the lensing properties of simulated cluster sized halos and
their possible use to constrain dark energy, see
e.g. \cite{meneghetti:2005b, meneghetti:2005a}. In the present study
we perform the first hydrodynamics simulation of cluster formation in
dark energy dominated universe and we focus, for the first time, on
the baryonic component of clusters and investigate the possible
effects of dark energy on their gas properties.  In an earlier study,
\cite{maio:2006} produced numerical simulations of dark energy with
baryonic gas to study the implications on cosmic reionisation from
first stars.

In all the aforementioned studies, the scalar field associated with
dark energy is assumed not to have density fluctuations on scales of
galaxy clusters or below. If dark energy influences the perturbations
on small scales as proposed for example by \cite{Arbey:2001qi},
\cite{Bean:2002kx}, \cite{Padmanabhan:2002sh} or \cite{Bagla:2002yn},
the collapse of structures itself will be affected. In our
simulations, we will also ignore any possibility for dark energy to
cluster and influence the cluster formation.

In addition to the cosmological constant model with a constant
equation of state parameter $w=\rho_{\rm de}/P_{\rm de}=-1$, we
simulate four other models previously studied in \cite{nunes2006}. 
These phenomenological models span the range of values that the
equation of state parameter can take for typical quintessence models
and are compactible with current observational constraints.
We take two models for which the dark energy is given by a perfect fluid
with constant equation of state parameter: $w = -0.8$ and $w=-1.2$
(phantom dark energy), and one model where dark energy results from a
slowly evolving scalar field in a potential with two exponential terms
(2EXP1)(\cite{barreiro2000})
\begin{equation}
V(\phi) = V_0 \left( e^{\alpha \kappa \phi} + e^{\beta \kappa \phi}\right) \,,
\end{equation}
where $\alpha = 6.2$ and $\beta = 0.1$ which has equation of state
parameter today $w_{0} = -0.95$. One other model (2EXP2) with varying
equation of state was studied in \cite{nunes2006}. However, because
its energy contribution to the total energy of the universe quickly
decays with redshift it has no significant departures from the Lambda
model in the redshift range of cluster formation (e.g. $z<5$). We thus
choose, from this point onwards, not to consider this model it in this
work.  
We further
assume for all models that energy density of dark energy, dark matter
and baryons today are respectively $\Omega_{\rm de} = 0.7$,
$\Omega_{{\rm m}}=0.3$, $\Omega_{{\rm B}}=0.0486$ and the Hubble
parameter $h=H_0/100 \, {\rm km\,s^{-1}}$ ${\rm Mpc^{-1}}=0.7$. 
$\sigma_8=0.9$.

\subsection{The simulation code}
 
For each of the cosmological models considered, we performed
$N$-body/hydrodynamic simulations of structure formation using a
modified version of the public {\tt Hydra} code \cite{couchman:1995,
pearce:1997}, which implements an adaptive
particle-particle/particle-mesh (AP$^3$M) algorithm to calculate
gravitational forces \cite{couchman:1991} and smoothed particle
hydrodynamics (SPH) \cite{monaghan:1992} to estimate hydrodynamic
quantities. The SPH implementation follows that used by
\cite{thacker2000} and conserves both energy and entropy. All
simulations included a model for radiative cooling using the method
described in \cite{thomas:1992} and based on the cooling tables of
\cite{sutherland:1993}. The metalicity was assumed to be a global time
varying quantity, $Z=0.3(t/t_0)Z_\odot$, where $t/t_0$ is the age of
the universe in units of the current time and $Z_\odot$ is the solar
metalicity. At a given time step, gas particles with temperatures
below $1.2\times 10^4$K and overdensities (relative to the critical)
larger than $10^4$, are converted into collisionless material and no
longer participate in the gas dynamical processes.

We modified the computation of the physical
time, $t$, and scale factor, $a$, in {\tt Hydra} to account for the effect of
time-variable equations of state of dark energy, 
\begin{equation}
t=\int_0^{a} {{{da'}\over{a'\,H(a')}}}=
H_0^{-1}\int_0^a {{{d\ln a'}\over {E(a')}}}\,,
\label{ageeq}
\end{equation}
where $E(a) \equiv H(a)/H_0$, or
\begin{equation}
\label{friedeq4}
E(a)^2={ {\Omega _m} \over {a^3}} +\Omega_{\rm de}\,e^{-3
\int_{a}^{1}{ 
(1+w(a')) 
{{da'}\over{a'}} 
}
}
+{ {1-\Omega _m-\Omega _{\rm de}}\over {a^2}} \,,
\label{Ezeq}
\end{equation} 
for flat cosmologies. To speed up computations, this quantity is
pre-tabulated for each of the dark energy models and read once at the
beginning of the simulation run. $E(a)$ is then interpolated and used
in Eq.~(\ref{ageeq}).  
With these modifications our version of the Hydra code can thus be
used to simulate generic models of homogeneous dark energy.

The initial density field of simulations was constructed, at redshift
$z=49$, using $N=4,096,000$ particles of baryonic and dark matter,
perturbed from a regular grid of fixed comoving size $L=100 \, h^{-1}
{\rm Mpc}$. We use the Zel'dovich approximation and the same set of
random numbers to generate the initial displacements. The amplitude of
the matter power spectrum was calculated assuming $\sigma_8=0.9$ in
all models and the linear growth factor computed for each model, as
presented in \cite{nunes2004}. We refer the reader to this article for
a discussion on the normalisation. We also assume that the matter
power spectrum transfer function is the same for all models and equals
that of the cosmological constant model, which is computed using the
BBKS formula \cite{bardeen:1986} and the shape parameter $\Gamma $
given by the formula in \cite{sugiyama:1995}. With this choice of
parameters, the dark matter and baryon particle masses are $2.1\times
10^{10} \, h^{-1} {\rm M_{\odot}}$ and $2.6 \times 10^{9} \, h^{-1}
{\rm M_{\odot}}$ respectively. In physical units the gravitational
softening was set fixed to $25\,h^{-1} {\rm kpc}$ below $z=1$ and
above this redshift scaled as $50(1+z)^{-1}\,h^{-1} {\rm kpc}$.
  
Individual simulation runs took between 2592 to 2812 time steps to
evolve to $z=0$. For each run we stored a total of 50 simulation
snapshots (box outputs) at a list of selected redshifts (the same to
all runs) ranging from $z=20$ to $z=0$. Thirty of these outputs are
inside the interval $0<z<3$, which is typically the range where galaxy
clusters form.

\begin{figure*}
\begin{center} 
\epsfxsize=8.30cm \epsfbox{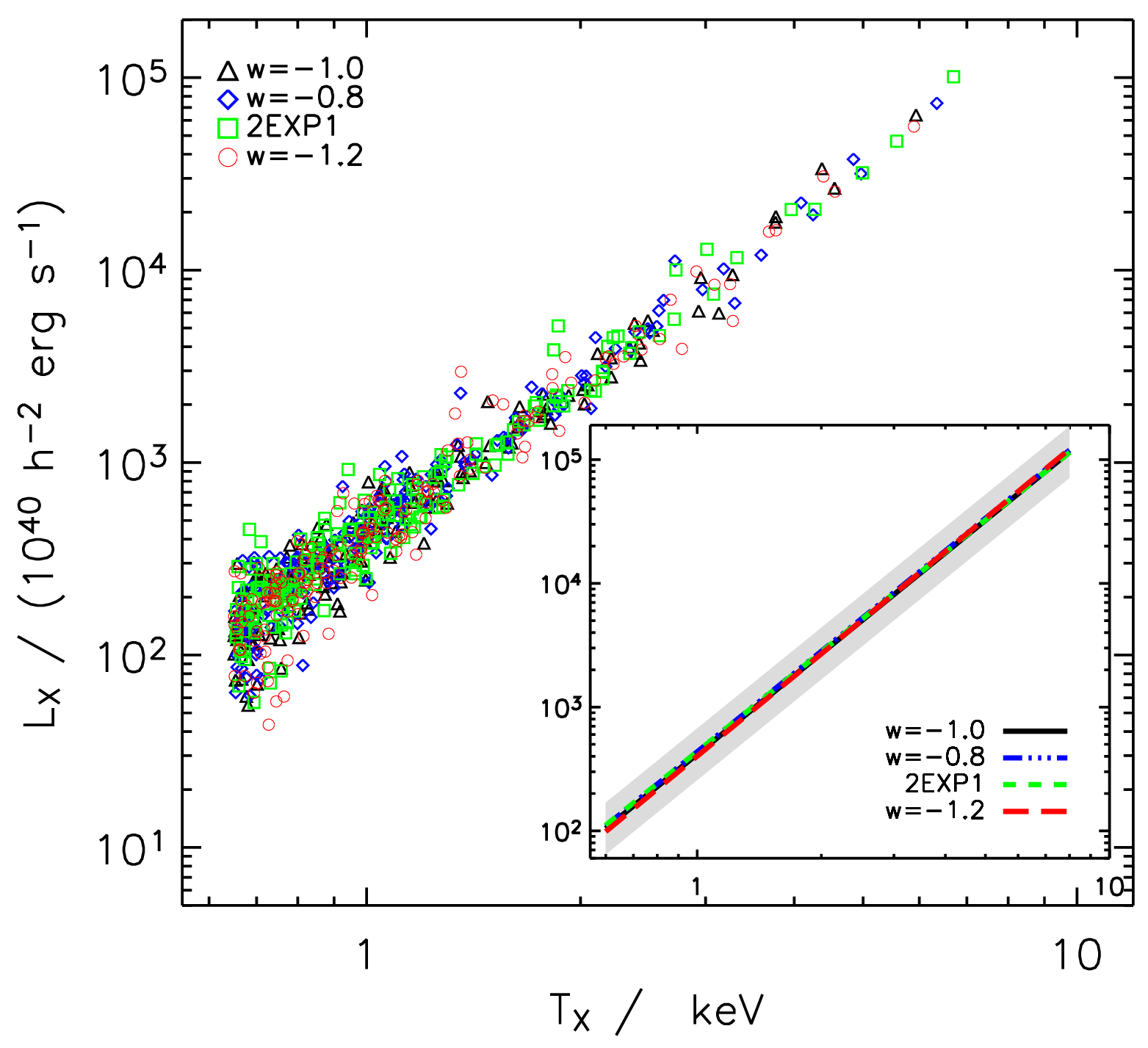}
\hspace{0.8cm}
\epsfxsize=8.30cm \epsfbox{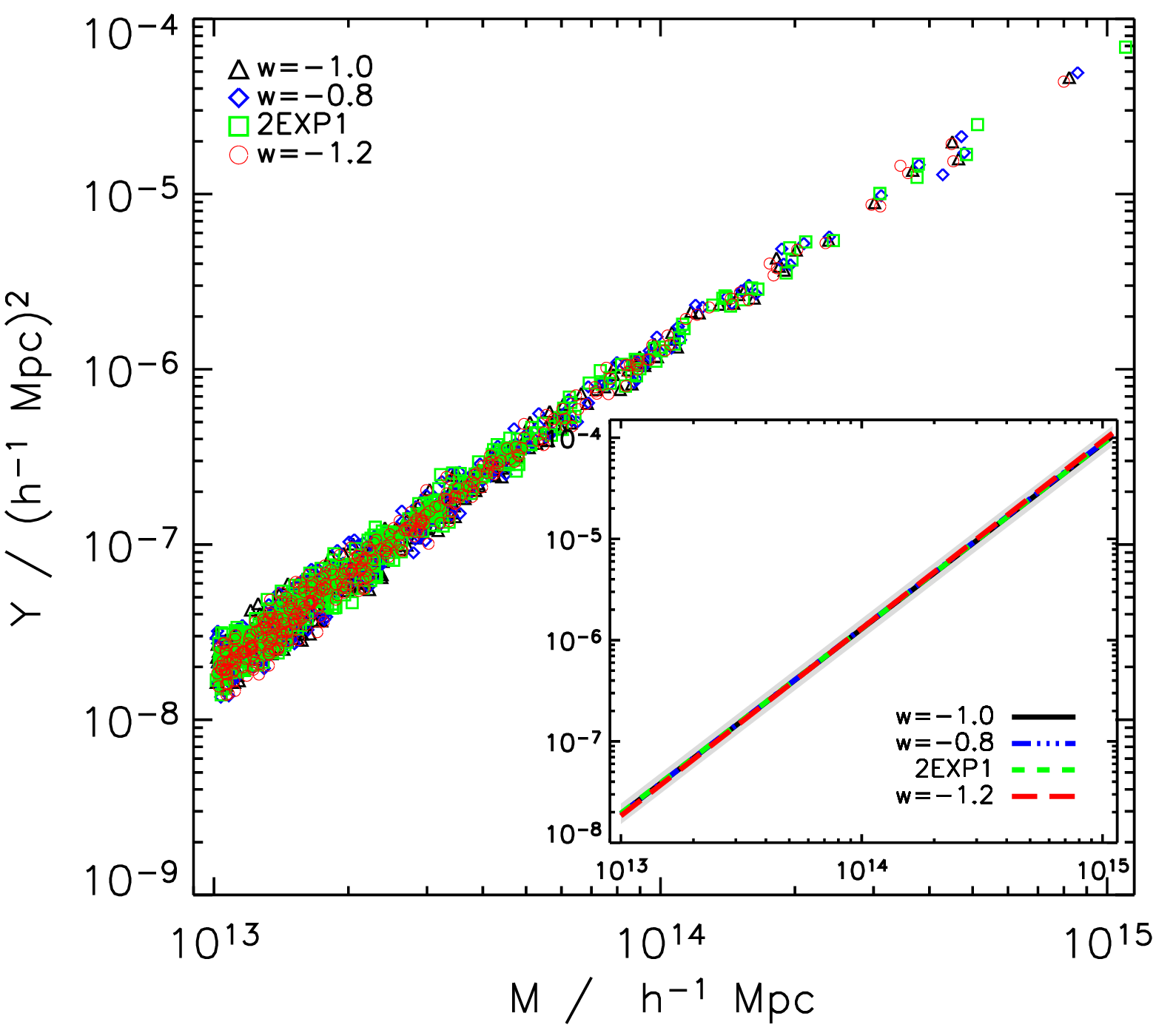}
\caption{\label{fig:scalings_z=0} Cluster scaling relations $L_{\rm
X,200}-T_{\rm X,200}$ (left panel) and $Y_{200}-M_{200}$ (right panel)
at redshift zero. Displayed quantities are computed within $R_{\rm
200}$, the radius where the mean cluster density is 200 times larger
than the critical density. The embedded plots show the best fits with
a power law to clusters represented in the main plots for the $w=-1$
(triangles), $w=-0.8$ (diamonds), 2EXP1 (squares) and $w=-1.2$
(circles) models. The shaded regions in the embedded plots give the
typical scatter of the fits, i.e. the r.m.s dispersion around the best
fit lines.}
\end{center}
\end{figure*}

\subsection{Catalogue construction} 
\label{catalogue}

From simulations, we constructed cluster catalogues using a modified
version of the cluster extraction software developed at Sussex by
Thomas and collaborators \cite{thomas:1998, pearce:2000,
muanwong:2001}. To summarize, the cluster identification process
starts with the construction of a minimal-spanning tree of dark matter
particles whose density exceeds the mean density of the box by
$\Delta_{\rm b}=178\times \Omega_m^{-0.55}(z)$ (i.e., the density
contrast predicted by the spherical collapse model of a virialised
sphere relative to the mean background density in the Lambda cosmology
\cite{eke:1998}). Although $\Delta_{\rm b}$ may differ for different
dark-energy models, this is not important at this step
because cluster properties are computed at fixed overdensities
as described below.
The minimal-spanning tree is then split into clumps of particles using
a maximum linking length equal to $0.5\,\Delta_{\rm b}^{-1/3}$ times the
mean inter-particle separation. Finally we grow a sphere around the
densest dark matter particle in each clump until the enclosed mass
verifies
\begin{equation}
M_\Delta(<R_\Delta)=\frac{4\pi}{3}R^3_\Delta\,\Delta\,\rho_{\rm crit}(z),
\label{eq:mass}
\end{equation}
where $\Delta$ is a fixed overdensity contrast, $\rho_{\rm
crit}(z)=(3H_0^2/8\pi G)E^2(z)$ is the critical density and
$E(z)$ is given by Eq.~(\ref{Ezeq}). We have constructed
master catalogues for all dark-energy simulation containing at least
500 particles of gas and dark matter, i.e. with an equivalent minimum
mass of $M_{\rm lim}\approx1.18\times10^{13}h^{-1} M_\odot$, at four
fixed overdensities, $\Delta=200, 500, 1000, 2500$. Here we will
report our findings for cluster scaling relations only at
$\Delta=200$, the largest cluster overdensity radius usually considered
in the literature. For this catalogue we find 377, 393, 396, 374
clusters at $z=0$ in the cosmological constant, $w=-0.8$, 2EXP1 and
$w=-1.2$ simulation runs respectively. Although our choice of
$\Delta_{\rm b}$ may limit exact comparison of numbers, these
abundances reproduce the behaviour predicted in \cite{nunes2006} in
their analytical study. We note that the cluster definition used in
this paper is different from that used in \cite{dasilva2004,
muanwong2006}.
Despite the similar simulation
parameters, direct comparison with their findings is only possible
for the Lambda model at redshift zero, where cluster definitions
are identical.
 
For each model, the cluster catalogue provides us with estimated
structural and observable quantities. More specifically for this
study we compute: masses; intrinsic SZ luminosity, $Y=Y_{\rm
SZ}\times D_{\rm A}^2$ where $Y_{\rm SZ}$ is the integrated SZ signal
and $D_{\rm A}$ is the angular diameter distance; mass-weighted
gas temperature $T_{\rm mw}$; and bolometric X-ray temperature,
$T_{\rm X}$, and luminosity, $L_{\rm X}$, excluding a cooling radius of
about 50 $h^{-1}$kpc around the cluster centre. We refer the reader to
\cite{dasilva2004} for the definitions of the these quantities.

\section{Analysis of the scaling relations}\label{sec:scal}

In this paper we investigate the following scaling relations between
cluster properties: $T_{\rm X}-M$, $Y-M$, $Y-T_{\rm mw}$, $L_{\rm
X}-T_{\rm X}$, and $Y-L_{\rm X}$. These can be expressed as:
\begin{equation}
T_{\rm X}=A_{\rm TM}\,(M/M_0)^{\alpha_{\rm TM}}\,(1+z)^{\beta_{\rm TM}}\, E(z)^{2/3}  \,,
\label{eq:tm}
\end{equation}
\begin{equation}
Y=A_{\rm YM}\,(M/M_0)^{\alpha_{\rm YM}}\,(1+z)^{\beta_{\rm YM}}\,E(z)^{2/3} \,,
\label{eq:ym}
\end{equation}
\begin{equation}
Y=A_{\rm YT}\,(T_{\rm mw}/T_{\rm mw,0})^{\alpha_{\rm YT}}\,(1+z)^{\beta_{\rm YT}}\, E(z)^{-1} \,,
\label{eq:yt}
\end{equation}
\begin{equation}
L_{\rm X}=A_{\rm LT}\,(T_{\rm X}/T_{\rm X,0})^{\alpha_{\rm LT}}\,(1+z)^{\beta_{\rm LT}}\, E(z) \,,
\label{eq:lt}
\end{equation}
\begin{equation}
Y=A_{\rm YL}\,(L_{\rm X}/L_{\rm X,0})^{\alpha_{\rm YL}}\,(1+z)^{\beta_{\rm YL}}\, E(z)^{-9/4} \,,
\label{eq:yl}
\end{equation}
where we have chosen the normalisation scales $M_0=10^{14}h^{-1}{\rm
M_{\odot}}$, $T_{\rm X,0}=T_{\rm mw,0}=1$ keV, $L_{\rm X,0}=10^{43}$
erg/s/h$^2$. The powers of the $E(z)$ function give the predicted
evolution, extrapolated from the self-similar model,
\cite{kaiser1986}, of the scalings in each case. The quantities, $A$,
$\alpha$, and $\beta$, give: the scalings normalisation at $z=0$; the
power on the independent variable; and the departures from the expected
redshift evolution. 

To investigate these cluster scaling relations in our simulations we
use the method described in \cite{dasilva2004}. 
According to Eqs.~(\ref{eq:tm}-\ref{eq:yl}), the
general form of how a given cluster property $y$ relates to a property
$x$ can be written as,
\begin{equation}
y\, f(z)=y_0(z) \,(x/x_0)^\alpha  \,, 
\label{eqyx}
\end{equation}
where 
\begin{equation}
y_0(z) = A \, (1+z)^\beta \,,
\label{eqy0z}
\end{equation}
and $f(z)$ is some fixed power of the cosmological factor
$E(z)$. This is a power-law function whose parameters $A$, $\alpha$
and $\beta$ can be obtained by fitting our cluster catalogue distributions at
each redshift with a straight line in the
$(\log(y\,f(z)),\log x)$ plane.
To be more specific, the fitting procedure is carried out in three
steps.  Firstly, we fit the cluster distributions with a straight-line
in logarithmic scale at all redshifts.
If the logarithmic slope $\alpha$ remains approximately constant (i.e. shows
no systematic variations) within the redshift range of interest, we then
take $\alpha$ at $z=0$ as the best fit value.
In the second step, we repeat the fit using this value of $\alpha$ to
determine the scaling normalisation factors $y_0(z)$. This avoids
unwanted correlations between $\alpha$ and $y_0(z)$.  Finally,
in the last step we use Eq.~(\ref{eqy0z}) to obtain the parameters
$A$ and $\beta$.

In the fitting process we consider only clusters with $M_{\rm lim}>5\times
10^{13}h^{-1}{\rm M_{\odot}}$ for scalings with mass, $L_{\rm lim} >
6.6 \times10^{42}h^{-2}$ erg s$^{-1}$ for scalings with luminosity, and $T_{\rm
mw,lim}> 1$ keV, $T_{\rm X,lim}> 1.1$ keV for scalings with mass-weighted and
emission-weighted temperatures, respectively. This ensures that our
cluster samples are complete at all redshifts and that, for each model,
equal number of clusters are used for all scalings at redshift
zero. With these selection criteria our cluster samples contain 60
clusters for the cosmological constant model, and a similar
number of clusters for the other models, at z=0.
We note that above $z=1.5$ the number of clusters with $M_{\rm lim}>5\times
10^{13}h^{-1}{\rm M_{\odot}}$ in our sample
decreases typically below 10, hence, we do not fit the scaling
relations above this redshift value. As we will discuss in the next
section, all the scaling relations explored in the present study, are
well fitted by power-laws of the form Eq.~(\ref{eqyx}), in the
redshift range $0<z<1.5$, except for the $L_{\rm X}-T_{\rm X}$ and
$Y-L_{\rm X}$ whose normalization dependences with redshift,
$\log(y_0(z))$, are well approximated by a straight line only in a
narrower redshift range.

\section{Results}\label{sec:res}

\subsection{Scaling relations at $z=0$}

We start by discussing the cluster scaling laws at redshift zero.
These were determined, for all scalings and models under investigation
in this work, at an overdensity radius $R_{\rm 200}$.  In Table
\ref{tabela} we present the power law best fit for the logarithmic
slopes $\alpha$ obtained for each of the cases.  As can be inferred
from the table, all models provide very similar results for each
scaling, with differences between models being comparable or in most
cases within the statistical errors of the fit. This indicates that
the local cluster scaling relations are quite insensitive to the
underlying dark energy model driving the present-day evolution of the
universe. 
As stated in Sec.~\ref{catalogue}, our cluster definition is the same
as in \cite{dasilva2004,muanwong2006} (only) at redshift zero, for the
Lambda model. We verified that indeed theirs and our results are in
excellent agreement at this redshift.

To illustrate the robustness of the scalings with respect to the dark
energy models investigated in this paper, we present in
Figure~\ref{fig:scalings_z=0} two characteristic X-ray and SZ galaxy
cluster scaling relations: the $L_{\rm X}-T_{\rm X}$ (left panel) and
the $Y-M$ (right panel) scalings.  In each case, the main plot shows
the cluster distributions for all models whereas the embedded panels
show the power law best fits obtained. For both scaling relations, the
cluster distributions and best fit lines clearly overlap. Also
represented by a shaded area in the embedded panels is the
r.m.s. dispersion of the fit for Lambda model:
\begin{equation}
\sigma_{\log y'} =\sqrt{ {1 \over N} \sum_{i} (\log (y'_i/y'))^2} \,,
\label{sigmay}
\end{equation}
where $y'=yf$ (see Eq.~(\ref{eqyx})) and $y'_i$ are individual data
points.
This dispersion is of the same size of the fit dispersions obtained in the
other models
and it is clearly wider than the best fit
line separations of the various dark energy models. As expected, the
scatter in the (core excised) $L_{\rm X}-T_{\rm X}$ is larger
than in the $Y-M$ relation due to the higher sensitivity
of the former scaling to the gas physics in the inner regions of
clusters.

\begin{figure}
\begin{center}
\epsfxsize=8.70cm \epsfbox{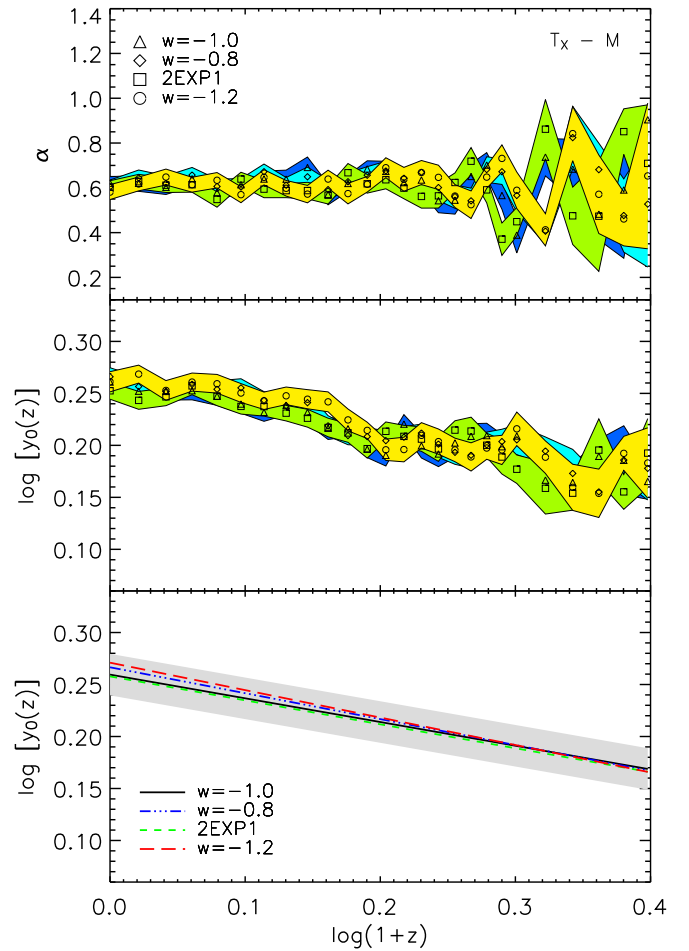}
\caption{\label{fig:tm} Slope (upper panel), normalisation
  $\log(y_0(z))$ (middle panel), and normalisation best fit lines
  (bottom panel) of the $T_{\rm X}-M$ relation as a function of
  redshift ($\log(y_0(z))$ is defined in Eq. \ref{eqy0z}). Blue colour
  and triangles stand for the cosmological constant model, cyan and
  diamonds are for the $w=-0.8$ model, green and squares are for the
  2EXP1 model, and yellow and circles are for the $w=-1.2$ model. The
  shaded area in the bottom panel gives the dispersion of the
  normalisation fit for the cosmological constant model.}
\end{center}
\end{figure}

\begin{figure}
\begin{center}
\epsfxsize=8.70cm \epsfbox{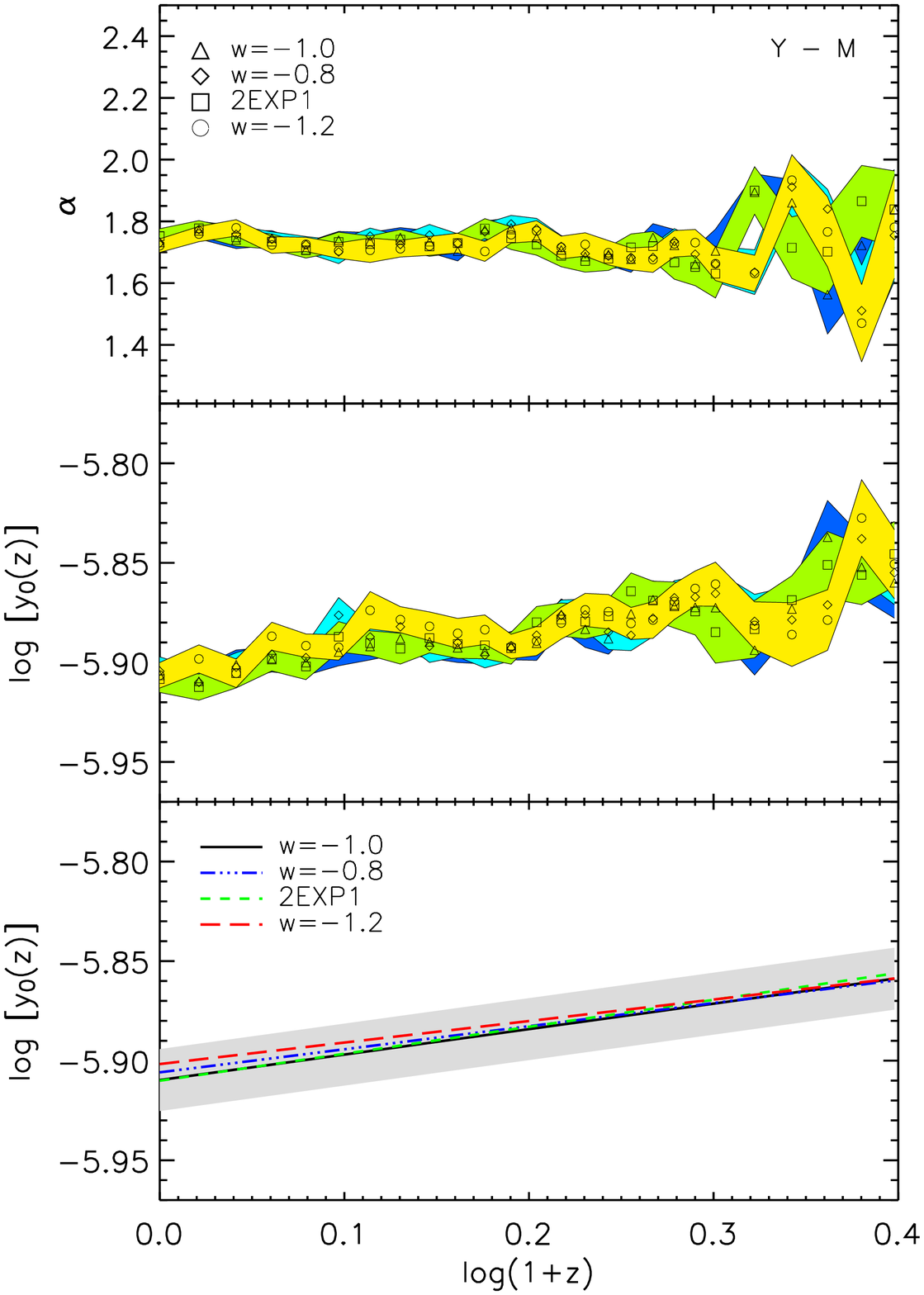}
\caption{\label{fig:ym} Slope, normalisation, and normalisation best
  fit lines of the $Y-M$ relation as a function of redshift. The
  shaded area in the bottom panel stands for the dispersion of the
  normalisation for the Lambda model.}
\end{center}
\end{figure}

\begin{figure}
\begin{center}
\epsfxsize=8.70cm \epsfbox{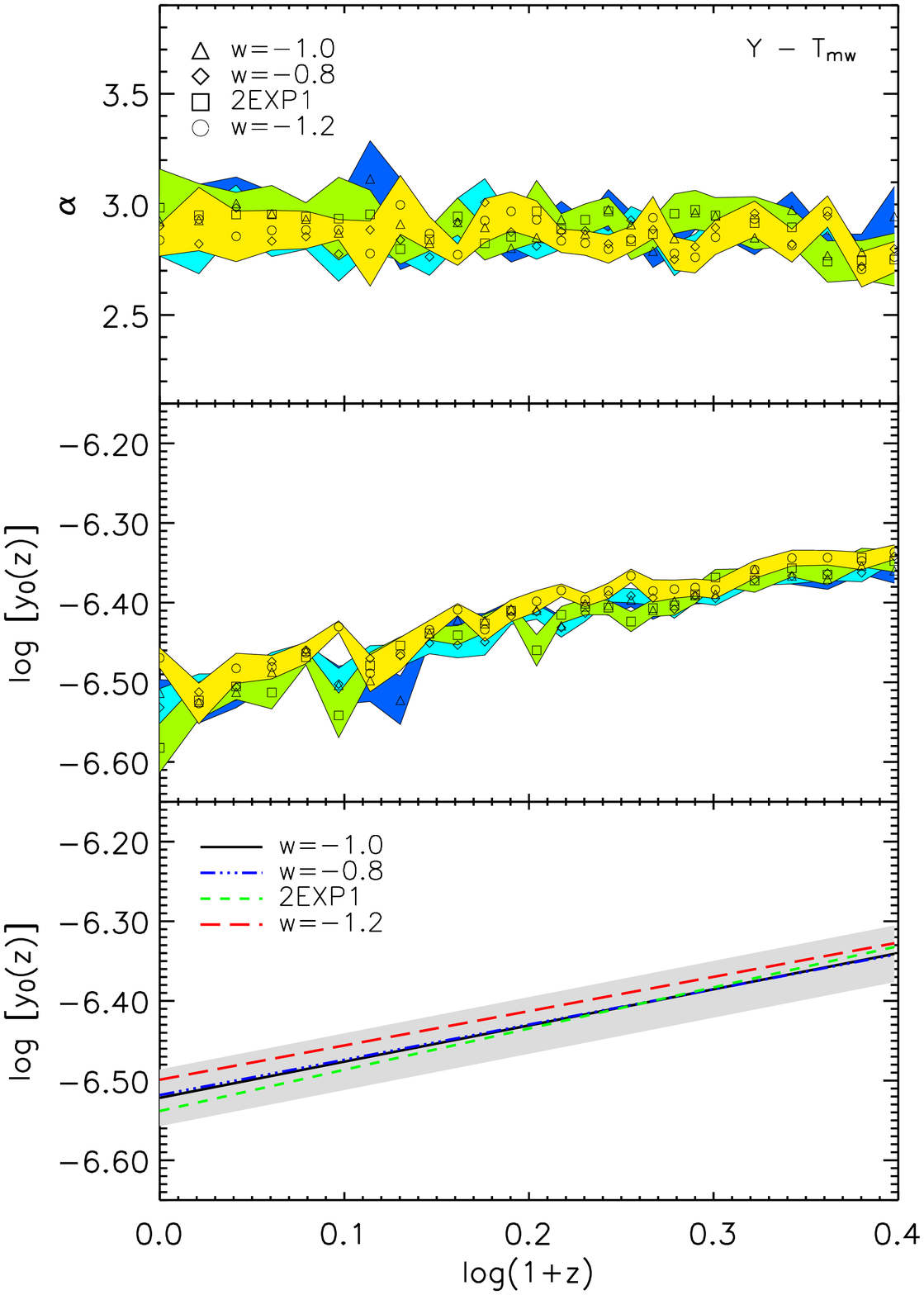}
\caption{\label{fig:yt} Slope, normalisation, and normalisation best
  fit lines of the $Y-T_{\rm mw}$ relation as a function of redshift. The
  shaded area in the bottom panel stands for the dispersion of the
  normalisation for the Lambda model.
}
\end{center}
\end{figure}

\begin{figure}
\begin{center}
\epsfxsize=8.70cm \epsfbox{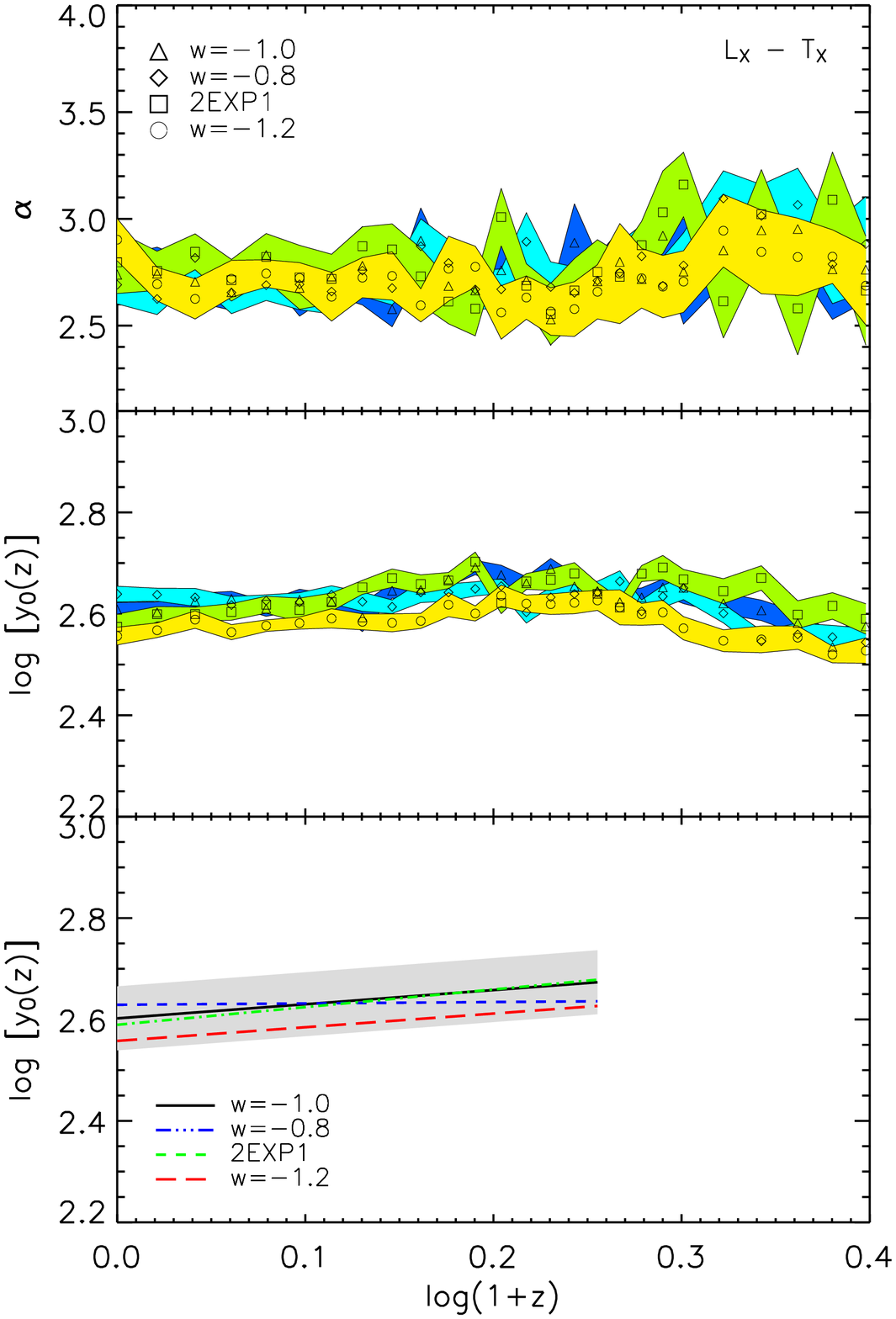}
\caption{\label{fig:lt} Slope, normalisation, and normalisation best
  fit lines of the $L_{\rm X}-T_{\rm X}$ relation as a function of redshift. The
  shaded area in the bottom panel stands for the dispersion of the
  normalisation for the Lambda model.
}
\end{center}
\end{figure} 
 
\begin{figure}
\begin{center}
\epsfxsize=8.70cm \epsfbox{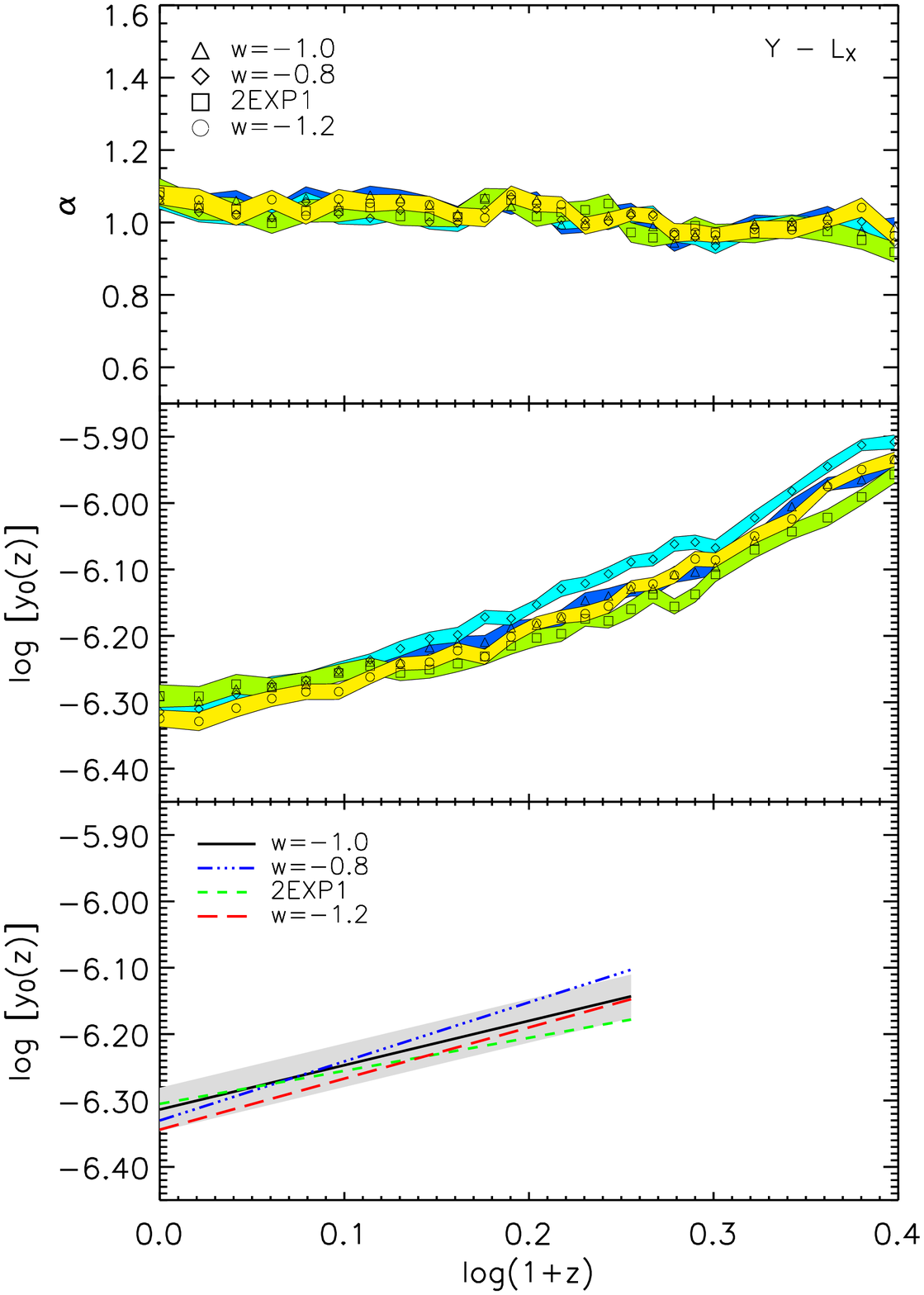}
\caption{\label{fig:yl} Slope, normalisation, and normalisation best
  fit lines of the $Y-L_{\rm X}$ relation as a function of redshift. The
  shaded area in the bottom panel stands for the dispersion of the
  normalisation for the Lambda model.
}
\end{center}
\end{figure}
 
\subsection{Evolution of the scaling relations}

In Figs.~\ref{fig:tm}, \ref{fig:ym}, \ref{fig:yt}, \ref{fig:lt} and
\ref{fig:yl} we present our findings for the variation with redshift
of $\alpha$ (top panels) and $y_0(z)$ (middle panels) for the $T_{\rm
X}-M$, $Y-M$, $Y-T_{\rm mw}$, $L_{\rm X}-T_{\rm X}$ and $Y-L_{\rm X}$
relations, respectively. 
Here, triangles, diamonds, squares and
circles represent the $w=-1$, $w=-0.8$, 2EXP1, and
$w=-1.2$ models, respectively. The coloured bands give the the best fit
errors obtained at each redshift for these quantities. 
The lines in the bottom panels of these figures are linear best fits
to the evolution of $\log (y_0(z))$ with $\log (1+z)$. The shaded area
in these panels gives the rms dispersion of the $\log (y_0(z))$ fit for
the cosmological constant model (similar values of dispersion are
found for the other models). The resulting best fit parameters for,
$A$, $\beta $ and $\alpha$ are presented in Table~\ref{tabela}, for
all scaling investigated in this paper.

As can be inferred from the figures, the fit to a power law, in a
redshift range $0<z<1.5$, was possible for all the scaling relations
explored in the present study except the relations involving $L_{\rm
X}$ (namely $L_{\rm X}-T_{\rm X}$ and $Y-L_{\rm X}$). For the latter
two, we found no significant departures of the slopes $\alpha$ from
the cosmological constant model but we could not fit the evolution of
the scaling relations by a power law within the whole redshift range
$0<z<1.5$ (see Figs. \ref{fig:lt} - \ref{fig:yl}). For these scalings we
therefore restrict the analysis to the a smaller range, namely
$0<z<0.75$, ie ($\log(1+z)<0.25$) where the linear fit is valid.

We compared the average value of the slope $\alpha$ of each scaling
relation, for all the cosmological models, over the redshift range
$0<z<1.5$ and found no significant departures from one model to the
other, confirming the behavior observed at $z=0$. This is also clear
from the upper panels of Figs.~\ref{fig:tm}, \ref{fig:ym},
\ref{fig:yt}, \ref{fig:lt} and \ref{fig:yl}, where there's a high
degree of overlap between the $\alpha$'s obtained for the four
cosmological models. 
For scalings with mass ($T_{\rm x}-M$ and $Y-M$) the degree of overlap
at high redshift is somewhat less striking due to larger variations
caused by the rapid decrease of the number clusters with $M_{\rm
lim}>5\times 10^{13}$. These ``oscillations'' show however no systematic
dependences with redshift and are about the same ``mean values'' for all
models.

In order to study the evolution of the scaling relations, we factored
out the redshift dependence expected from the self-similar evolution
which is parametrised by a power law with exponent $\beta$, see
Eq.~(\ref{eqy0z}). 
As a general remark, we find that, excluding the $T_{\rm x}-M$
relation, all scalings investigated here show positive evolutions
relative to the expected self-similar evolution (i.e. for a given $x$
in Eq~(\ref{eqyx}) the property $yf$ is higher at higher redshift). In
the case of $T_{\rm x}-M$ relation the opposite behaviour is
observed. This is expected because the inclusion in all models of
radiative cooling (which is a non-gravitational physical process)
causes cluster scaling laws to deviate from self-similar evolution,
see eg \cite{dasilva2004,muanwong2006} for studies in cosmological
constant model simulations.

It is clear from the table and the lower panels of the figures that
the value of $\beta$ is slightly more model dependent than the
parameters $\alpha$ and $A$. This is especially the case for the
$L_{\rm X}-T_{\rm X}$, and $Y-L_{\rm X}$ relations for
which there is a mild difference between models. For these scalings
the w=-1.2 and w=-0.8 model generally show the largest deviations from
the cosmological constant model, whereas the 2EXP1 show the smallest.
However, those differences are of the same order as the
intrinsic errors and dispersions. We can thus safely consider that there
are no significant departures from the cosmological constant model.

\section{Conclusions}\label{sec:conc}

The abundance of clusters, their redshift distribution, as well as
their clustering, are governed by the geometry of the universe and the
power spectrum of the initial density perturbations. Gas physics
related to cluster structure and evolution also enters through mapping
of the cluster observable (SZ flux or X-ray luminosity) relative to
the total mass of the cluster. As a result, galaxy cluster counts can
be used as probes of cosmological and cluster properties. However,
there are several requirements needed to achieve precise cosmological
constraints: (i) advances in understanding the formation and evolution
of cluster size halos; (ii) a good understanding of the selection
function; (iii) robust observational proxies for the cluster mass.

The first condition relates to the conduction of large
simulations. This is greatly achieved in the standard cosmological
model with a cosmological constant. In the context of dark energy
dominated universe, N-body simulations are becoming available for
models different from the simple cosmological constant with constant
or varying equation of state parameter. These simulations now provide us with a
good understanding of the halo properties. They show that the
halo mass function is well approximated by the Jenkins
mass function. Simulations also indicate that the clusters halos are more
concentrated in dark energy models, since structure grow earlier, than
in lambda models and that concentrations are higher in models with
varying than constant equation state. 
The second condition,
understanding the cluster selection function, translates in
understanding the limiting mass and the completeness of the surveys
from realistic mock cluster catalogues. The mass (or
temperature) selection function is directly linked with the cluster
observed quantities through cluster scaling relations which is our
third requirement (the need for a good proxy for the cluster mass).

In this work we have thus for the first time explored the scaling laws
for both SZ and X-rays observations using hydrodynamic simulations of
galaxy clusters in four dark energy models with constant or varying
equation of states spanning a large class of models. We have studied
the scaling properties at $z=0$ and their evolution with redshift. We
have found that dark energy induces no modifications on the scaling
laws at $z=0$ and presents very little differences from the
cosmological constant model at higher redshifts.

While detailed simulations incorporating viable dark energy models
remains a program in progress, it is reassuring that all models
considered in this work predict similar scaling properties to the
lambda model. The modeling of the cluster gas component appears to be
nearly independent of the dark energy model.  Therefore, using the
``standard'' lambda model scaling relations for converting observable
to masses and temperature in future surveys should not introduce any
additional bias in the cosmological constraints derived from cluster
counts. In this work, we have considered that dark energy does not
cluster with dark matter. It would be interesting, however, to
evaluate how our conclusions stand for numerical simulations in a
scenario where dark energy is inhomogeneous and collapses along with
dark matter during the formation of structure. This will be pursued in
a forthcoming study.

\begin{table*}
\begin{center}
\caption{\label{tabela} Best fit values of the parameters $\alpha$,
  $\log A$ and $\beta$ as well as their respective
  $1\sigma$ errors. These values are valid within the redshift range
  $0<z<1.5$.  For the $L_{\rm X}-T_{\rm X}$ and $Y-L_{\rm X}$
  relations, the linear fit and the associated parameters are only
  valid in the range $0<z<0.75$ above which the linear fit is not a
  good approximation.}
\begin{tabular}{|l|l|rrrr|}
\hline
Model    &     & $w=-1$ & $w=-0.8$ & 2EXP1 & $w=-1.2$  \\
\hline
$T_{\rm X}-M$  & $\alpha_{\rm TM}$  & $0.620\pm 0.029$ & $0.604\pm
0.031$ & $0.581\pm 0.033$ & $0.602\pm 0.029$\\
& $\log A_{\rm TM}$ & $0.260\pm 0.005$ & $0.267\pm 0.004$ & $0.271\pm 0.005$ & $0.258\pm 0.005$ \\
  & $\beta_{\rm TM}$ & $-0.228\pm 0.020$ & $-0.249\pm 0.017$ & $-0.264\pm 0.023$ & $-0.230\pm 0.023$  \\
\hline
$Y-M$  & $\alpha_{\rm YM}$ & $1.732\pm 0.025$ & $1.730\pm 0.025$ &
$1.721\pm 0.022$ & $1.752\pm 0.024$ \\
& $\log A_{\rm YM}$ & $-5.910\pm 0.004$ & $-5.906\pm 0.004$ & $-5.902\pm 0.005$ & $-5.910\pm 0.003$ \\
 & $\beta_{\rm YM}$  & $0.128\pm 0.016$ & $0.116\pm 0.016$ & $0.108\pm 0.020$ & $0.135\pm 0.013$  \\
\hline
$Y-T_{\rm mw}$ & $\alpha_{\rm YT}$ & $2.922\pm 0.100$ & $2.902\pm
0.136$ & $2.838\pm 0.072$ & $2.985\pm 0.175 $\\
& $\log A_{\rm YT}$ & $-6.522\pm 0.008$ & $-6.518\pm 0.005$ & $-6.499\pm 0.007$ & $-6.538\pm 0.008$\\
 & $\beta_{\rm YT}$  & $0.454\pm 0.036$ & $0.443\pm 0.022$ & $0.430\pm 0.031$ & $0.517\pm 0.036$\\
\hline
$L_{\rm X}-T_{\rm X}$  & $\alpha_{\rm LT}$ & $2.738\pm 0.086$ &
$2.691\pm 0.089$ & $2.902\pm 0.099$ & $2.796\pm 0.146$ \\
& $\log A_{\rm LT}$ & $2.602\pm 0.010$ & $2.629\pm 0.007$ & $2.558\pm
0.006$ & $2.589\pm 0.011$ \\
& $\beta_{\rm LT}$ & $0.279\pm 0.063$ & $0.027\pm 0.042$ & $0.270\pm 0.035$ & $0.348\pm 0.070$  \\
\hline
$Y-L_{\rm X}$  & $\alpha_{\rm YL}$ &  $1.063\pm 0.028$ & $1.064\pm
0.026$ & $1.076\pm 0.026$ & $1.084\pm 0.037$ \\
& $\log A_{\rm YL}$ & $-6.314\pm 0.005$ & $-6.330\pm 0.004$ & $-6.344\pm 0.005$ & $-6.305\pm 0.006$ \\
& $\beta_{\rm YL}$  & $0.668\pm 0.033$ & $0.890\pm 0.028$ & $0.770\pm 0.034$ & $0.497\pm 0.037$ \\
\hline
\end{tabular}
\end{center}
\end{table*}

\begin{acknowledgements}

The authors are indebted to Peter Thomas, Orrarujee Muanwong and
collaborators for the their part in writing the original Sussex cluster
extraction software used in this work. We thank Orrarujee Muanwong for
discussions and comments on the manuscript. We also acknowledge PAI-PESSOA
collaboration program as well as a partial support from the CNES and
Programme National de Cosmologie. The simulations used in this study
were performed at the IAS computing facilities. NA thanks CAUP for
hospitality. AdS acknowledges support from Funda\c c\~ao Ci\^encia e
Tecnologia (FCT) under the contracts SFRH/BPD/20583/2004 and CI\^ENCIA
2007.

\end{acknowledgements}


\begin{thebibliography}{}
\bibitem[Allen \& Fabian 1998]{allen98}
Allen, S.W. \& Fabian, A.C. 1998, \mnras, 297, L57
\bibitem[Arbey et al. 2001]{Arbey:2001qi}
{Arbey}, {A.}, {Lesgourgues}, J. {Salati}, P. {2001}, {Phys. Rev. D},
{64}, {123528}.
\bibitem[Arnaud et al. 2005]{arnaud2005}
Arnaud, M., Pointecouteau, E., Pratt, G.W., 2005, Astron \&
Astrophys., 441, 893. 
\bibitem[Ascasibar et al. 2006]{ascasibar2006}
Ascasibar, Y., Sevilla, R., Yepes, G., Mueller, V., Gottloeber,
S. 2006, M.N.R.A.S., 371, 193. 
\bibitem[Babul et al 2002]{babul2002}
Babul, A., Balogh, M.L., Lewis, G.F., Poole, G.B. 2002, \mnras, 330, 329.
\bibitem[Bagla et al. 2003]{Bagla:2002yn}
{Bagla}, {J.~S.}, {Jassal}, {H.~K.}, {Padmanabhan} ,.
  {2003}, {Phys. Rev.}, {D67}, {063504}.
\bibitem[Balogh et al. 2006]{balogh2006}
Balogh, M.L., Babul, A., Voit, M., McCarthy, I.G., Jones, L.R., Lewis,
G.F., Ebeling, H. 2006, M.N.R.A.S., 366, 624. 
\bibitem[Bardeen et al. 1986]{bardeen:1986} Bardeen, J.~M., Bond,
J.~R., Kaiser, N., \& Szalay, A.~S., 1986, \apj, 304, 15
\bibitem[Barreiro et al. 2000]{barreiro2000}
{Barreiro}, {T.}, {E.~J.} {Copeland}, and {N.~J.} {Nunes},
{2000}, {Phys. Rev. D}, {61}, {127301}.
\bibitem[Bartelmann et al. 2005]{bartelmann05}
Bartelmann, M., Doran, M., Wetterich, C. 2005, astro-ph/0507257
\bibitem[Battye \& Weller 2003]{battye2003} {Battye}, {R.~A.}, and
{J.}~{Weller}, {2003}, {Phys. Rev.}, {D68}, {083506}.
\bibitem[Bean \& Magueijo 2002]{Bean:2002kx}
{Bean}, {R.}, \& {Magueijo}, J. {2002}, {Phys. Rev.},
 {D66}, {063505}.
\bibitem[Benson et al. 2004]{benson2004}
Benson, B.A., Ade, P.A.R., Bock, J.J.
\bibitem[Bialek et al. 2001]{bialek2001}
Bialek, J.J., Evrard, A.E., Mohr, J.J. 2001, \apj, 555, 597.
\bibitem[Bonaldi et al. 2007]{bonaldi2007}
Bonaldi, A., Tormen, G., Dolag, K., Moscardini, L. 2007,
astro-ph/0704.2535. 
\bibitem[Borgani et al. 2004]{borgani2004}
Borgani, S., et al. 2004, \mnras, 348, 1078.
\bibitem[Bryan \& Norman 1998]{bryan98}
Bryan, G.L. \& Norman, M.L. 1998, \apj, 495, 80.
\bibitem[Cooray 1999]{cooray99}
Cooray, A.R. 1999, \mnras, 307, 841.
\bibitem[Couchman 1991]{couchman:1991} Couchman, H.~M.~P., 1991, Ap.J,
368, L23
\bibitem[Couchman et al. 1995]{couchman:1995} Couchman H. M. P.,
        Thomas P. A., Pearce F. R., 1995, \mnras, 452, 797
\bibitem[Dolag et al. 2004]{dolag:2004} Dolag, K., Bartelmann, M.,
Perrotta, F., Baccigalupi, C., Moscardini, L., Meneghetti, M., Tormen,
G., 2004, \mnras, 416, 853 
\bibitem[Edge \& Stewart 1991]{edge91}
Edge, A.C. \& Stewart, G.C. 1991, \mnras, 252, 414.
\bibitem[Eke Navarro \& Frenk 1998]{eke:1998}
Eke, V.~R., Navarro, J.~ F., \& Frenk, C.~S., 1998, \apj, {\bf 503}, 569
\bibitem[Ettori et al. 2004]{ettori2004} 
Ettori, S., Tozzi, P., Borgani, S., Rosatti, P. 2004, Astron. \&
Astrophys., 417, 13  
\bibitem[Evrard et al. 1996]{evrard96}
Evrard, A.E., Metzler, C.A. \& Navarro, J.F. 1996, \apj, 469, 494.
\bibitem[Finoguenov et al. 2001]{finoguenov2001}
Finoguenov, A., Reiprich, T.H., Bohringer, H. 2001, Astron. \& Astrophys., 
368, 749.
\bibitem[Haiman et al. 2001]{haiman2001}
{Haiman}, {Z.}, {J.~J.} {Mohr}, and {G.~P.} {Holder},
{2001}, {Astrophys. J} {553}, {545}.
\bibitem[Henry 2004]{henry2004}
Henry, J.P. 2004, \apj, 609, 603. 
\bibitem[Jenkins et al. 2001]{jenkins:2001} Jenkins, A., Frenk,
C.~S., White, S.~D.~M., Colberg, J.~M., Cole, S., Evrard, A.~E.,
Couchman, H.~M.~P., Yoshida, N., 2001, \mnras, 321, 372
\bibitem[Kaiser 1986]{kaiser1986} Kaiser N.,
        1986, MNRAS, 222, 323
\bibitem[Kay et al. 2007]{kay2007}
Kay, S.T., da Silva, A.C., Aghanim, N., Blanchard, A., Liddle, A.R.
Puget, J.-L., Sadat, R., Thomas, P.A. 2007, MNRAS,
\bibitem[Klypin et al.2003]{Klypin:2003ug}
{Klypin}, {A.}, {Maccio}, {A.~V.}, {Mainini}, R., {Bonometto},
{S.~A.} {2003}, \apj, {599}, {31}.
\bibitem[Kuhlen et al. 2004]{Kuhlen:2004rw}
{Kuhlen}, {M.}, {Strigari}, {L.~E.}, {Zentner}, {A.~R.}, 
{Bullock}, {J.~S.},  {Primack}, {J.~R.} {2004}, {astro-ph/0402210}.
\bibitem[Linder \& Jenkins 2003]{Linder:2003dr}
{Linder}, {E.~V.} \& {Jenkins}, J. {2003}, {astro-ph/0305286}.
\bibitem[Lokas et al. 2003]{Lokas:2003cj}
{Lokas}, {E.~L.}, {Bode}, P., {Hoffman}, Y. {2003}, {astro-ph/0309485}.
\bibitem[Maio et al. 2006]{maio:2006} Maio, U., Dolag, K.,
Meneghetti, M., Moscardini, L., Yoshida, N., Baccigalupi, C.,
Bartelmann, M., Perrotta, F., 2006,\mnras,373, 869
\bibitem[Manera \& Mota 2006]{manera2006}
Manera, M., Mota, D. F. 2006 \mnras, 371, 1373.
\bibitem[Markevitch 1998]{markevitch98}
Markevitch, M. 1998, \apj, 504, 27.
\bibitem[Maughan et al. 2006]{maughan2006}
Maughan, B.J., Jones, L.R., Ebeling, H., Scharf, C. 2006, \mnras, 365,
509. 
\bibitem[McCarthy et al. 2003]{mccarthy2003}
McCarthy, I.G., Babul, A., Holder, G.P., Balogh, M.L. 2003, \apj, 591, 515.
\bibitem[Meneghetti et al. 2005a]{meneghetti:2005a}Meneghetti, M., Jain, B.,
Bartelmann, M., Dolag, K., 2005, \mnras, 362, 1301
\bibitem[Meneghetti et al. 2005b]{meneghetti:2005b} Meneghetti, M.,
Bartelmann, M., Dolag, K., Perrotta, F., Baccigalupi, C., Moscardini,
L., Tormen, G., 2005, Astron. \& Astrophys., 442, 413
\bibitem[Mohr 2004]{mohr2004}
{Mohr}, {J.~J.}, {2004}, {astro-ph/0408484}.
\bibitem[Molt et al. 2005]{molt2005}
Molt, P.M., Hallman, E.J., Burns, J.O., Norman, M.L. 2005, \apj, 623,
L63.  
\bibitem[Monaghan 1992]{monaghan:1992}
Monaghan, J.~J., 1992, Ann. Rev. Astron. Astrophys., 30, 543 
\bibitem[Morandi et al. 2007]{morandi2007}
Morandi, A., Ettori, S., Moscardini, L. 2007 {astro-ph/0704.2678}.
\bibitem[Muanwong et al. 2001]{muanwong:2001}
Muanwong, O., Thomas, P.A. Kay, S.T., Pearce, F.R., Couchman,
H.M.P. 2001, \apj, 552, L27. 
\bibitem[Muanwong et al. 2002]{muanwong:2002} Muanwong 
        O., Thomas P. A., Kay S. T., Pearce F. R., 2002, MNRAS, 336, 527
\bibitem[Muanwong et al. 2006]{muanwong2006}
Muanwong, O., Kay, S.T., Thomas, P.A. 2006, \apj, 649, 640.
\bibitem[Nevalainen et al. 2000]{nevalainen2000}
Nevalainen, J, Markevitch, M., Forman, W. 2000, \apj, 532, 694.
\bibitem[Nunes \& Mota 2006]{nunes2004}
{Nunes}, {N.~J.}, and {D.~F.} {Mota}, {2006}, M.N.R.A.S., 368, 751.
\bibitem[Nunes et al. 2006]{nunes2006}
{Nunes}, {N.~J.}, da {Silva}, A.C., Aghanim, N.
{2006}, Astron. \& Astrophys., 450, 899.
\bibitem[Padmanabhan \& Choudhury 2002]{Padmanabhan:2002sh}
{Padmanabhan}, {T.} \& {Choudhury}, T.R. {2002}, Phys. Rev.,
  {D66}, {081301}.
\bibitem[Pearce \& Couchman 1997] {pearce:1997} 
        Pearce F. R., Couchman H. M. P., 1997, New Astronomy, 2, 411
\bibitem[Pearce et al 2000] {pearce:2000} Pearce F. R., Thomas P. A., 
Couchman H. M. P., Edge A. C., 2000, MNRAS, 317, 1029 
\bibitem[Perlmutter et al. 1999]{perlmutter1999}
Perlmutter, S. et al. 1999, \apj, 517, 565.
\bibitem[Rasia et al. 2005]{rasia2005}
Rasia, E., Mazzotta, P., Borgani, S., Moscardini, L., Dolag, K., 
Tormen, G., Diaferio, A., Murante, G. 2005, \apj, 618, L1.
\bibitem[Riess et al. 1998]{riess1998}
Riess, A. G. et al. 1998, \aj, 116, 1009.
\bibitem[Rowley et al. 2004]{rowley2004}
Rowley, D.R., Thomas, P.A. \& Kay, S.T. 2004, \mnras, 352, 508.
\bibitem[da Silva et al. 2000]{dasilva2000} da Silva A. C., Barbosa D.,
         Liddle A. R., Thomas P. A., 2000, MNRAS, 317, 37
\bibitem[da Silva et al. 2001]{dasilva2001} da Silva A. C., Barbosa D.,
         Liddle A. R., Thomas P. A., 2001, MNRAS, 326, 155
\bibitem[da Silva et al. 2004]{dasilva2004}
da Silva, A.C., Kay, S.T., Liddle, A.R., Thomas, P.A. 2004,
M.N.R.A.S., 348, 1401
\bibitem[Sugiyama 1995]{sugiyama:1995}
Sugiyama, N., 1995, Astrophys.J.Suppl., 100, 281
\bibitem[Sunyaev \& Zel'dovich 1972]{sunyaev72} 
Sunyaev R. A., Zel'dovich  Ya. B., 1972, Comm. Astrophys. Space Phys., 4, 173
\bibitem[Sunyaev \& Zel'dovich 1980]{sunyaev80}
Sunyaev R. A., Zel'dovich Ya. B., 1980, ARA\&A, 18, 537 
\bibitem[Sutherland \& Dopita 1993]{sutherland:1993}
Sutherland R. S., Dopita M. A., 1993, \apjs, 88, 253
\bibitem[Thacker \& Couchman 2000]{thacker2000} 
Thacker R. J., Couchman H. M. P., 2000, \apj, 545, 728
\bibitem[Thomas \& Couchman 1992]{thomas:1992}
 	Thomas P. A., Couchman H. M. P., 1992, MNRAS, 257, 11
\bibitem[Thomas et al. 1998]{thomas:1998} Thomas P. A. 
        et al. (the Virgo Consortium), 1998, MNRAS, 296, 1061
\bibitem[Thomas et al. 2001]{thomas2001}
Thomas, P.A., Muanwong, O., Pearce, F.R., Couchman, H.M.P., Edge, A.C., 
Jenkins, A., Onuora, L. 2001, \mnras, 324, 450
\bibitem[Voit et al. 2002]{voit2002}
Voit, G.M., Bryan, G.L., Balogh, M.L., Bower, R.G. 2002,
\apj, 576, 601. 
\bibitem[Wang et al. 2004]{wang2004}
{Wang}, {S.}, {J.}~{Khoury}, {Z.}~{Haiman}, and {M.}~{May}, 2004,
{Phys. Rev. D} 70, {123008}.
\bibitem[Weinberg \& Kamionkowski 2003]{weinberg2003}
{Weinberg}, {N.~N.}, \&
{M.}~ {Kamionkowski}, 2003, M.N.R.A.S., 341, 251.
\bibitem[Weller et al. 2001]{weller2001}
{Weller}, {J.}, {R.}~{Battye}, and {R.}~{Kneissl},
{2002}, {Phys. Rev. Lett.}, {88}, {231301}.

\end{thebibliography}
\end{document}